\documentclass[prb,twocolumn,showpacs]{revtex4-1}

\usepackage{amsmath}
\usepackage{amssymb}
\usepackage{amsfonts}
\usepackage{dsfont}
\usepackage{enumerate}
\usepackage{bm}
\usepackage{graphicx}
\usepackage{wrapfig}
\usepackage{setspace}
\usepackage{color}
\usepackage{cancel}
\usepackage{verbatim}

\def\tr{\mbox{tr}}

\renewcommand{\>}{\rangle}
\newcommand{\<}{\langle}

\def\XXint#1#2#3{{\setbox0=\hbox{$#1{#2#3}{\int}$}
     \vcenter{\hbox{$#2#3$}}\kern-.5\wd0}}

\begin{document}

\title{Integrable Quantum Hydrodynamics in Two Dimensional Phase Space}
\author{E. Bettelheim}
\address{Racah Inst. of Physics, Hebrew University of Jerusalem, 91904 Israel}

\begin{abstract}
Quantum liquids in two dimensions represent interesting dynamical quantum  systems for several reasons, among them the possibility of the existence of infinite hidden symmetries, such as conformal symmetry or the symmetry associated with area preserving diffeo-morphisms. It is known that when the symmetry algebra is large enough, symmetry may fully prescribe the dynamics. However, the way this is borne out in two dimensional hydrodynamics, both classical and quantum, is not fully understood. Here we take a step in clarifying this issue, by focusing on a particular example, namely that of a two dimensional phase space liquid which emerges when one considers the Calogero model, a many-body one-dimensional system interacting through an inverse square law potential. We demonstrate how the symmetry algebra of  conserved quantities of the one dimensional system is expressed in terms of the incompressible Euler hydrodynamics of point vortices in phase space. Due to a formal relation between quantum hydrodynamics and classical stochastic hydrodynamics, which is inherent in the method of stochastic quantization, the ideas and methods developed here may also have application in the study of stochastic classical hydrodynamics, after suitable modifications.
\end{abstract}

\maketitle

\section{Introduction}
Hidden symmetries are a fascinating aspect of solvable systems . Indeed, solvable models are often only so because of the existence of enough conserved quantities to pin down the solution entirely\cite{Takhtajan:Bethe:Ansatz:Review,Sutherland:Book:Beautiful:Models}. An important example of a symmetry algebra that is often large enough to  afford a full solution is conformal symmetry\cite{Belavin:Polyakov:Zamolodchikov,ginsparg:Applied:CFT}. Conformal symmetry has been suggested long ago as an underlying symmetry for 2D turbulence by Polyakov \cite{Polyakov:Turbulence}. Until this date, it is not clear how to apply Polyakov's original idea to turbulent flows. Nevertheless, more recently it was shown that conformal invariance in two dimensional turbulence may be observed in the fractal geometry of zero vorticity isolines \cite{Falkovich:Bernard:CFT:Turbulence} It should be noted, though, that Polyakov was concerned with the direct cascade while the latter observation was made in the inverse cascade.

One may then say that it remains an open question  how the infinite dimensional symmetry of conformal invariance appears within two dimensional stochastic hydrodynamic systems. In order to make progress towards answering this question, we pose the following, slightly different question: how do infinite symmetries (including conformal invariance) appear in {\it quantum} two dimensional hydrodynamic systems?

Our stated goal is then to find a model which on the one hand exhibits quantum hydrodynamic flows and on the other hand exhibits an infinite symmetry algebra and to provide the tools to further study the model both from the hydrodynamics perspective and from the perspective of the representation theory of the infinite algebra, especially the conformal one\cite{Feigin:Fuchs:Review:Rep:CFT}.

Our strategy will be to start with a model well known to possess infinite symmetry and map it to a hydrodynamic problem. The model in question  is a one dimensional system of particles on a line with inverse square potential and a harmonic trap, namely, the Calogero model\cite{Calogero,Ha:Book,Perelomov:Olshanetsky:Calogero:Projection,Perelomov:1971} and the hydrodynamics are that of point vortices.
We then use ideas developed by   Susskind\cite{104:Susskind:NC} and Polychronakos\cite{Polychronakos:FQH:CS} to represent the phase space dynamics of such a fluid by Euler dynamics of an incompressible fluid. Such methods are also familiar in classical hydrodynamic problems of free boundaries \cite{Zakharov1968} We then study the symmetries of the fluid by considering the known symmetry\cite{Wadati:QCalogero:And:W} algebra of the Calogero-Sutherland model and by translating those symmetries to  hydrodynamical language.

In order to achieve this we make use of a quantization scheme of the Calgero-Sutherland model, which takes into account that this model is described by a Hamiltonian reduction of a matrix harmonic oscillator. The quantization scheme we use is that of stochastic quantization. The latter has the advantage that such a quantization is especially lucid for Hamiltonian reductions, as it takes into account constraints and gauge conditions in an intuitive way\cite{Faddeev:Constriants:Singular:Lagrangians} (the alternative being using Faddeev-Popov ghosts, or a quantization scheme which holds for only certain values of  the interaction strength\cite{Polychronakos:FQH:CS} ). At the same time the stochastic quantization approach may be  useful in  discerning the common features, and indeed the differences, between  quantum hydrodynamics and classical stochastic hydrodynamics. This last theme,  however, lies beyond the scope of this paper, and will be the subject of future work.

Our approach uses different formal tools, but  is closely related  to the quantum hydrodynamics approach\cite{Landau:1941:Q:Hydrodynamics}, which quantizes hydrodynamics starting from a classical Poisson bracket structure\cite{Volovik:Poisson:Brackets,Rasetti:Regge:Vortex:Algebra,Montgomery:Marsden:Hamiltonian:Structure:Free:Boundary}. This approach has been recently applied to the quantum fractional Hall effect\cite{Wiegmann:FQHE:Edge,Wiegmann:Kirchoff:Larking,Wiegmann:Kirchoff:Short,Abanov:FQHE:Hydrodynamics}, which again is a closely related subject\cite{Stone:Superfluid:FQHE:Hydrodynamics}. In the context of the Hall effect the relevance of infinite symmetries and in particular those related to area preserving diffeomorphisms has been discussed in [\onlinecite{76:Cappelli:Wsymmetry}].
  
\section{Calogero Model as Matrix Model}
As mentioned, we shall be studying the Calgero model, namely a set of $N$ particles on the\ infinite  line with a confining harmonic potential. We assume, as standard,   that the center of mass is fixed at the origin $ \sum_{i=1}^N x_i  = 0.$ The  Hamiltonian for this model is: \begin{align}\label{CalgeroH}
H= \sum_i \left( \frac{ p_i^2}{2m} + \frac{m \Omega^2 \ x_i^2}{2}\right) + \sum_{i\neq j}\frac{\theta^2}{m}\frac{1}{\left( x_i - x_j\right)^2} .
\end{align}
There are two length scales in the problem, $\ell_\theta$ and $\ell_\hbar$, which will be used extensively  below. These are given by:
\begin{align}
\ell_\theta= \sqrt{\frac{\theta}{m\Omega}}, \quad \ell_\hbar= \sqrt{\frac{\hbar}{m\Omega}}.
\end{align}

Central to our approach, is the fact that this model can be recast in the terms of  the dynamics of matrices,  an observation of Perelomov \cite{Perelomov:1971,Perelomov:Olshanetsky:Calogero:Projection}. Consider then two  Hermitian traceless matrices ,  $\hat X$ and $\hat Y$, which have the following commutation relation,
\begin{align}\label{constraintXY}
\left[\hat X,\hat Y\right]=\imath\ell_\theta^2\left[\mathds{1}-(N+1)|v\>\<v|\right],
\end{align}
for some unit vector $|v\>$. Postulate the following, quite trivial, dynamics for the matrices:
\begin{align}
\dot{\hat X} =\Omega \hat Y, \quad \dot{\hat Y}= - \Omega \hat X.\label{evolutionXY}
\end{align}

These evolution equations  are easy to solve. The solution being   $\hat X(t) = \hat X(0) \cos(\Omega t) + \hat Y(0)\sin (\Omega t )$ and $\hat Y(t) = \hat Y(0)\cos(\Omega t) - \hat X(0)\sin (\Omega t ).$ Despite this simple form for the time dependence of the matrices, the eigenvalues perform  complicated dynamics. Indeed, the eigenvalues of the matrix $\hat X(t)$ if multiplied by $\sqrt{m \Omega}$ obey the dynamics prescribed by the Hamiltonian (\ref{CalgeroH}), with the diagonal elements of $\hat  Y(t)$ serving as momenta if those are divided by $ \sqrt{m \Omega}$.

It is easier to work with complex matrix coordinates by defining $\hat Z=\hat X + i\hat  Y$. The evolution equations, Eqs. (\ref{evolutionXY}), then take the form:
\begin{align}\label{evolutionZ}
\dot{\hat Z} = - \imath \Omega \hat Z,
\end{align}
where now  the constraint reads:
\begin{align}\label{constraintZ}
[\hat Z,\hat Z^\dagger] =  2 \ell_\theta^2\left[\mathds{1}-(N+1)|v\>\<v|\right.].
\end{align}

The relation between the matrix model (\ref{evolutionXY}, \ref{constraintXY}) (or equivalently the matrix model defined by (\ref{evolutionZ},\ref{constraintZ})) and the Calogero model is realized by examining the eigenvalues of the matrix $\hat X$ which are revealed  by applying unitary transformation $U^\dagger \hat X U = \Lambda$, where $\Lambda$ is a diagonal matrix, if $U|v\> = e^{\imath \beta} |v\>$,  the constraint is invariant to such a unitary transformation, acting both on $\hat X$ and $\hat Y$.  As a consequence, one may 'gauge' the matrix theory and write down the following dynamics (written for the matrix $Z$ for convenience):
\begin{align}\label{evolutionWithGauge}
\dot{\hat Z} = -\imath \Omega \hat Z + \imath [\lambda,\hat Z],
\end{align}
supplemented by the constraint (\ref{constraintZ}) and the condition
\begin{align}\label{lambdaCommutes}
[\lambda, |v\>\<v|]=0.
\end{align}

The function $\lambda$ may be used to realize a desired gauge. For example, to obtain the Calogero dynamics from the matrix model, the gauge in which $\hat X$ is diagonal may be employed. We may then assume:
\begin{align}
\<i|\hat X|j\> =  \sqrt {m \Omega} x_i \delta_{i,j}.
\end{align}
In this gauge one takes the vector $|v\>$ to be given by:
\begin{align}
\<i|v\> = \frac{1}{\sqrt{N}}
\end{align}
To satisfy the constraint, Eq. (\ref{constraintXY}), $Y$ takes the form:
\begin{align}
\<i|\hat Y|j\> = \frac{ p_i}{\sqrt{m\Omega}}\delta_{i,j} + (1-\delta_{i,j})\frac{\imath\theta}{\sqrt{m\Omega}}\frac{1}{x_i-x_j},
\end{align}
while the condition that $X$ remains diagonal implies that $\lambda$ must take the form:
\begin{align}
\<i|\lambda|j\>  =\frac{\theta}{m}\left(-(1-\delta_{i,j})\frac{1}{(x_i-x_j)^2}  + \delta_{i,j} \sum_{k\neq i}\frac{1}{(x_i-x_k)^2}\right),\nonumber
\end{align}
which is consistent with (\ref{lambdaCommutes}). The evolution equation, Eq. (\ref{evolutionWithGauge}), then implies the Calogero dynamics for $x_i$ and $p_i$:
\begin{align}\label{CalogeroDynamics}
\dot{x}_i = \frac{ p_i}{m}, \quad \dot{p}_i =\sum_{j \neq i} \frac{2\theta^{2}/m}{(x_i-x_j)^3}.
\end{align}

The procedure of starting from (\ref{evolutionZ}, \ref{constraintZ}) and ending up at the Hamiltonian dynamics prescribed by (\ref{CalgeroH}), namely, (\ref{CalogeroDynamics}), is an example \cite{Gonera:Calogero:Hamiltonian:Reduction} of a Hamiltonian reduction.

A Hamiltonian reduction is a procedure by which a system with a certain number of degrees of freedom (in our case the $2N^2-2$ real degrees of freedom associates with the two  traceless Hermitian matrices), is subjected to a certain number of constraints (here the matrix constraint (\ref{constraintZ}) actually imposes $N^2-N$ conditions), and the same number of gauge conditions (demanding that the matrix $X$ is diagonal indeed applies $N^2 - N  $ conditions) to obtain Hamiltonian dynamics on a smaller set of degrees of freedom (here the $2N-2$ degrees of freedom associated with $N$ coordinates $N$ momenta from which the degrees of freedom of the center of mass coordinate and the center of mass momentum must be subtracted as they are assumed to be fixed).

Denoting the constraint conditions as $\phi^{(a)} =0$ and the gauge conditions as $\chi^{(b)}=0$, one must apply the following requirements in order for these conditions to be deemed a Hamiltonian reduction. First the gauge conditions must Poisson-bracket commute:
\begin{align}\label{chichic0}
\{ \chi^{(a)} , \chi^{(b)}\} =0.
\end{align}
Second, the constraints must be irreducible. That is, there must be a symbol $d^{a,b}_c$ for which:
\begin{align}\label{PhiIrreducible}
\{\phi^{(a)},\phi^{(b)}\} =\sum_c d^{a,b}_c \phi^{(c)}.
\end{align}
And lastly there must be a non-singular matrix $\Delta^{a,b}$, such that:
\begin{align}
\{\phi^{(a)},\chi^{(b)} \}  = \Delta^{a,b}.
\end{align}
The fact that $\Delta^{a,b}$ is non-singular allows  the constraints, $\phi^{(a)}$ to generate all gauge transformations, and the gauge conditions $\chi^{(b)}$ may be applied to generate flows to enforce the constraint.

We shall not show here that the gauge conditions implied by demanding that the matrix $X$ is diagonal conform to the requirements of a Hamiltonian reduction, as this was done in [\onlinecite{Gonera:Calogero:Hamiltonian:Reduction}], and in fact we shall be interested in another gauge, in which the hydrodynamic interpretation becomes more directly apparent. We stress, though, that the two gauges are two different Hamiltonian reductions of the same system.

\section{Coulomb gauge and semiclassics}

The gauge we shall pursue was suggested by Susskind\cite{104:Susskind:NC}, and dubbed Coulomb gauge. The gauge condition reads:
\begin{align}\label{gauge}
[a,\hat Z^\dagger] + [a^\dagger , \hat Z ]=0,
\end{align}
where $a$ is an $N\times N$ matrix, obtained by projecting the standard annihilation operator:
\begin{align}
a = \sum_{j=1}^N \ell_\theta \sqrt{ j}|j-1\>\<j|\label{aDef}
\end{align}
Note that $a$ itself is a particular solution of the constraint (\ref{constraintZ}) and gauge condition (\ref{gauge}), due to the following fact:
\begin{align}
[a,a^\dagger ] =  2\ell_\theta^2\left[\mathds{1}-(N+1)|v\>\<v|\right],
\end{align}
which can be easily verified.

To motivate the particular gauge choice, (\ref{gauge}), we shall develop a semiclassical approach to the matrices in the large $N$ limit. To do so we shall associate with each operator its Wigner transform, $\hat{M} \to M(x,p)$, where:
\begin{align}\label{WignerT}
M(x,p) =   \sum_{i,j} \<i|\hat{M}|j\> \int \left\<i\right. \left|x+\frac{y}{2} \right\>\left\<x-\frac{y}{2} \right|j\> e^{\frac{i p y}{\theta}} dy .
\end{align}
Here $\<i|\hat{M}|j\>$ is simply the matrix element $\hat{M}_{i,j}$, while, in order to make sense of $\left\<i\right. \left|x+\frac{y}{2} \right\>$ and $\left\<x-\frac{y}{2} \right|j\>$, we may prescribe that $\{\<i|\}_{i=0}^\infty$ be any complete ortho-normal basis of the Hilbert space. We make the following choice, which turns out to be convenient:
\begin{align}
\<x|i\> =\frac{1}{\pi^{1/4} \sqrt{\ell_\theta 2^i i!}} H_i \left(\frac{x}{\ell_\theta}\right) e^{-\frac{x^2}{2\ell_\theta}}.
\end{align}

With eq. (\ref{WignerT}) we have the usual limit of an operator $\hat{M}$ as a function $M(x,p)$ of the phase space variables. 

We now discuss at the semiclassical limit of the matrices we have defined and the various matrix constraints  that are applied in the na\"{i}ve limit where the matrices become infinite. Namely, first the matrices size is sent to infinity and then the Wigner transform is computed, rather than computing the Wigner transform, then applying appropriate re-scaling procedures to ensure a good $N\to\infty$ limit and only then  sending the matrix size to infinity.  Indeed, if one formally send $N$ to infinity, one obtains that the matrix $a$ has a limit $w=x+i p $. Making use of the fact that the commutator becomes the Poisson bracket with respect to the variable $x$ and $p$, the gauge condition then reads:
\begin{align}\label{gaugeNableTimesR}
\vec{\nabla}_w \times \vec{R} (x,p)= 0,
\end{align}
while the constraint reads:
\begin{align}\label{XcommaYis1}
\{ X , Y \} = 1,
\end{align}
where $\vec{R}$ is the vector $\vec{R} (x,p)= \left(\begin{array}{c} X(x,p) \\ Y(x,p) \end{array}\right)$ and the subscript $w$ on $\vec{\nabla}_w$ denotes that the derivative is taken with respect to $x$ and $p$. The Poisson brackets in (\ref{XcommaYis1}) are taken with respect to the canonical variables $x$ and $p$, while $\hat X$ and $\hat Y$ are the Wigner tranforms of the matrices ${X}$ and $ Y$, namely, $\{X,Y\} = \frac{\partial X}{\partial x}\frac{\partial Y}{\partial p}-\frac{\partial Y}{\partial x}\frac{\partial X}{\partial p}$.  

To obtain a hydrodynamic interpretation of the evolution equations and constraints in the semiclassical representation of the $N\to\infty$ limit, we adopt Susskind's interpretation of the $w$ coordinate (or, equivalently, the $x$ and $p$ coordinate system) as being Lagrangian (material) coordinates of particles, while $Z(x,p) = X(x,p) + \imath Y(x,p),$ (or, equivalently, the $X(x,p)$ and $Y(x,p)$ coordinate system) as the Euler coordinates (the physical coordinates of the particles).

Eq. (\ref{XcommaYis1}) is the statement that the Jacobian of the transformation from the Lagrangian to the Eulerian frame is $1$. If we assign the constant density of particle in the Lagrangian frame to be $\rho_0 = \frac{1}{\ell_\theta^2}$, we obtain that the density in the physical, Eulerian, frame is constant and equal to $\rho_0 = \frac{1}{\ell_\theta^2}$.  This implies incompressible flows of the particles, as the density is constant. If we differentiate the gauge condition in time, we obtain that the vorticity must be 0, at least as viewed from the $w$ frame, and hence we obtain irrotational incompressible flows, which in the infinite plane with vanishing boundary conditions at infinity means that there are no flow at all. However, this is only a consequence of the crude way in which the $N\to\infty$ limit was taken. The true form of the semiclassics will be a refinement of this na\"{i}ve\ limit.

\section{Hydrodynamics in the Coulomb Gauge}

In order to achieve non-trivial hydrodynamics we shall proceed more carefully with the large $N$ and the semiclassical interpretation of the different quantities of interest. To obtain hydrodynamics, we first define the density:
\begin{align}
\rho(\vec{R}) =\left.  \frac{P(x,p)}{\{X,Y\}} \right|_{\vec{R}(x,p) = \vec{R}}\label{densityJ},
\end{align}
where $ \vec{R}(x,p) = \left( \begin{array}{c} X(x,p) \\ Y(x,p) \end{array}\right),  $ and $P(x,p)$ is the Wigner transform of the $N\times N$ identity matrix. This definition for the  density reflects a situation in which in the $x,p$ plane,  which in turn the density of particles is given  by $P(x,p)$, while that in the $\vec{R}$-plane is obtained by multiplying by the proper Jacobian of the  transformation  between the two planes, which is just one over the Poisson bracket between $X$\ and $Y$.

To motivate further the definition of the density consider that a trace of an operator, $\hat{O}$, may be written in a standard fashion through its Wigner transform $O(x,p)$  as:
\begin{align}
\tr(\hat O) = \int O(x,p) P(x,p) dx dp =\int O(\vec{R} )\rho(\vec{R}) d^2\vec{R}, \nonumber
\end{align}
where $O(\vec{R})$ is just $O(x,p)$ after a transformation of coordinates to the $\vec{R}$-plane. 

The actual functional form of the density may only be computed once $X$ and $Y$ are chosen to have a specific form. Eq. (\ref{XcommaYis1}) states that the na\"{i}ve expectation is for the density to be equal just to $P(x,p)$. The latter function can be computed explicitly, we do not give the details here, but it is not surprising that it turns out to be given by a function close to $1$ inside a circle of radius $\sqrt{2 N } \ell_\theta$, while close to zero outside.

It is important to note that the Poisson bracket $\{X,Y\}$  is a time independent function of $(x,p)$. Namely,  it is constant in a Lagrangian coordinate system. The reason for this is that the Poisson bracket is the semiclassical limit of the left hand side of the constraint, (\ref{constraintZ}), while the right hand side in (\ref{constraintZ}) is time independent.  Given the relation between the Poisson bracket we have that the density is time independent in Lagrangian coordinates, namely the flows are incompressible. 

Taking a time derivative of the density one obtains:
\begin{align}
\partial_t \rho + \vec{\nabla} (\dot{\vec{R}} \rho)=0.
\end{align}
To obtain hydrodynamic flows, we assume that the density obtained by applying (\ref{densityJ}) is close to $1$ in most of the space inside a circle of radius $\sqrt{2N } \ell_\theta $, with the exception of small regions in which it varies. We have the following approximate solution for the constraint in the region where the density is close to $1$:
\begin{align}\label{smallDeviation}
Z = a + \bar{V}(a^\dagger,t).
\end{align}
With this { \it\ ansatz}, the constraint is in fact satisfied to first order in $V$. The generator, $\lambda$ , which insures that the constraint remains being obeyed with the flow of time is given as follows:
\begin{align}
\lambda =-  \frac{\Omega}{\ell_\theta^2} a^\dagger a,
\end{align}
indeed the time evolution in this case is given by:
\begin{align}\label{ZdotFofV}
\dot{Z} = -i \Omega\left(  V(a^\dagger,t) + V'(a^\dagger) a^\dagger \right).
\end{align}
This last equation  shows that $Z$ retains its form, (\ref{smallDeviation}), in the evolution. In fact, to first order in $V$  taking the semiclassical limit  and applying the Wigner transform yields the following (the Poisson bracket having the same definition as in (\ref{XcommaYis1})):
\begin{align}
\dot{V} = -\imath \Omega V  -  \frac{\Omega}{\ell_\theta^2}  \{ Z^\dagger Z,V\}.
\end{align}
The last term on the right hand side represents a simple rotation in space. Moving to a rotating frame, changes the derivative $\frac{d}{dt} f\to \frac{d}{dt} f+   \frac{\Omega}{\ell_\theta^2}\{ Z^\dagger Z, f \}$, which gets rid of this term. Thus in the rotating frame we have:
\begin{align}
\dot{V} = -\imath \Omega V .
\end{align}
This trivial dynamics should not be taken too seriously, indeed it was arrived at by  dropping all higher terms in $V$. Nevertheless, it implies that in the regions where the density is close to $1$, and, consequently, $V$ is small, we have
\begin{align}\label{vIszcrossdeltaR}
\vec{v} = \Omega  \hat{z} \times \delta \vec{R},
\end{align}
where here $\delta \vec{R}$ is given by $V = \delta R_x + i \delta R_y $. 

Eq. (\ref{vIszcrossdeltaR}) suffices in recovering hydrodynamics, if it is supplemented by  incompressibility of the flows, which, we remind, is a consequence of the fact that the right hand side of  (\ref{densityJ}) is time independent.
To show that, we compute the total vorticity inside some contour, $\partial C$, which goes over the regions where the density is close to $1$. We have:
\begin{align}
&\int_C  \omega \ d^2\vec{R}  =  \ \oint_{\partial C} \vec{v} \cdot d\vec{l} = \\
&=\Omega \ \oint_{\partial C} \hat{z} \times \delta \vec{R} \cdot d\vec{l}  =\frac{\theta}{m } \int_C \delta \rho  d^2\vec{R}.
\end{align}
 Here we only use (\ref{vIszcrossdeltaR}) in the region where it is valid, namely in the region where the deviation, $\delta \vec{R}$, is small. The latter relation shows that $ \omega = \frac{\theta}{m} \delta \rho$. Since $\delta \rho$ satisfies a continuity equation, so does $\omega$. This explicitly reads:
\begin{align}\label{VorticityFlow}
\partial_t \omega + \vec{v} \cdot \vec{\nabla} \omega =0.
\end{align}
This last equation is in fact equivalent to Euler hydrodynamics;
\begin{align}\label{Euler}
\partial_t \vec{v} + \vec{v} \cdot \vec{\nabla}\vec{v}  + \vec{\nabla}P=0.
\end{align}
Note that pressure in incompressible 2D hydrodynamics is not an independent variable but may be extracted by knowledge of the velocity.

We stress that (\ref{VorticityFlow}) and (\ref{Euler}) are shown here to hold only as the effective equations for well separated vortices. The Calogero model phase space description was shown to have quantized vortex-type excitation with circulation quantum equal to $\Omega l_\hbar^2 $ in [\onlinecite{104:Susskind:NC},\onlinecite{Polychronakos:FQH:CS}].   
The full description of the dynamics, including motion within vorticity patches, namely the dynamics of a generic continuous vorticity field, is not discussed in this manuscript and we do not show that the dynamics has hydrodynamic form. Nonetheless, vortex dynamics is a rather rich subset of generic hydrodynamic flows.

\section{Stochastic Quantization }

We shall use stochastic quantization in phase space and apply gauge conditions and constraints to obtain a Hamiltonian reduction. Here we use a version of Fadeev's path integral  quantization as applied to system with Dirac type constraints \cite{Dirac:Generalized:Hamiltonian:1950,Dirac:Generalized:Hamiltonian:1958},  utilizing  stochastic quantization instead, in a method due to Ohba \cite{Ohba:StochQuanti:Constraints}.  However,  we leave out gauge and constraints in this section, describing only the general framework of stochastic quantization in phase space.

Stochastic quantization \cite{Parisi:Wu} postulates  motion in auxiliary time. We shall denote this time by $ {\tau}$ to distinguish it from physical time $ {t}$. In the phase space version, stochastic motion is then prescribed  such that at very large auxiliary time, $ {\tau}\to\infty$, any correlation functions  of a combination of phase space variables will converge to the quantum result.

Specifically,  given a Euclidean action $S_{E} $ of the form:
\begin{align}\label{SE}
S_{E}(\tau) \ = \int_{{t}_i}^{{t}_f} \left[ i \sum_i \dot{p_i} q_i -H\left(\{p_i\}_{i=1}^N,\{q_i\}_{i=1}^N\right)\right]d{t},
\end{align}
where the right hand side is to be evaluated at a fixed auxiliary time, $ \tau$, the following stochastic motion in the phase space  may be postulated:
\begin{align}\label{qEuclideanStochaticQ}
\frac{dq_i({t},{\tau})}{d   \tau } = -\frac{\delta S_{E}}{\delta q_i({t},{\tau})} + d\xi^{(1)}_i( {t}, {\tau}),\\ \quad \label{pEuclideanStochaticQ}\frac{dp_i( {t}, {\tau})}{d \tau} = -\frac{\delta S_{E}}{\delta p_i( {t}, {\tau})}+ d \xi^{(2)}_i( {t}, {\tau})
\end{align}
where
\begin{align}
\< d\xi_i^{(a)} ( {t}, {\tau}) \ d\xi_j^{(b)}( {t}, {\tau})\>=2\hbar \delta_{i,j} \delta _{a,b} \delta ( {\tau}- {\tau}')\delta ( {t}- {t}').
\end{align}

Quantum mechanics is recovered at $ \tau \to \infty$, at which point a steady state distribution is reached for the phase space variables. Steady state means that $\tilde q_i( {t}, {\tau})$ and $p_i( {t}, {\tau})$ do not change in distribution as a function of $ {\tau}$.

Note that because of the term $i\dot{p}q $ in the Euclidean Lagrangian, the stochastic motion in (\ref{qEuclideanStochaticQ},\ref{pEuclideanStochaticQ}), does not translate into a simple classical stochastic motion in phase space.  This interpretation is obstructed by the imaginary unit, which requires  complexifying phase space.  Nevertheless, the two problems (quantum vs. classical stochastic) are related formally in the following sense. Classical stochastic motion may be recovered by considering special solutions of (\ref{qEuclideanStochaticQ}, \ref{pEuclideanStochaticQ}) in which the $\tau$ derivatives vanish rather than vanishing only in distribution. In addition, physical time  must be Wick rotated to the imaginary axis. To make use of this formal relation between the two problems, one must have analytical methods to solve the stochastic motion, which is the case where the quantum problem is integrable. Fleshing out this relation is beyond the scope of the current paper.

\section{The Coulomb gauge as a Hamiltonian Reduction}

 An alternative approach to quantizing the matrix model (\ref{evolutionZ})  is to augment the symmetry of the problem by relaxing condition (\ref{lambdaCommutes}), which amounts to promoting the gauge symmetry from $SU(N)\times U(1)$ to $SU(N+1)$. This, however, restricts to values of the coupling constant $\frac{\theta}{\hbar}$ to be $\frac{1}{n}$. The version of stochastic quantization that we shall employ may be carried through without any restriction, at least formally. We do not address the question of the convergence of the stochastic process we shall define, as we are interested in connecting the stochastic motion to symmetries, which are likely not to be sensitive to the question of convergence.

We start by explicitly writing the  constraint in the form $\tilde\phi^{i,j}=0$:
\begin{align}
\tilde\phi^{i,j} = 2 \imath \<j| [\hat X,\hat Y]|i\> - \ell_\theta^2 (\delta_{i,j} -N \delta_{i,N} \delta_{j,N}).
\end{align}
The gauge may be written in the form $\tilde\chi^{i,j}=0$ for :
\begin{align}
\tilde\chi^{i,j} =2\imath \left([\hat{x},\hat X]+[\hat{y},\hat Y]\right) _{i,j},
\end{align}
where $\hat{x} = \frac{\hat{a} + \hat{a^\dagger}}{2}$ and  $\hat{y} = \frac{\hat{a} - \hat{a^\dagger}}{2 \imath}$.

The following may be computed directly:
\begin{align}\label{phiphi}
&\left\{ \tilde\phi^{i,j} , \tilde\phi^{k,l} \right\} = 2 \imath \left( \delta_{i,l}\tilde\phi^{k,j}-\delta_{j,k} \tilde\phi^{i,l} \right) + \\
&+ \nonumber 4 \imath N \theta (\delta_{i,N} \delta_{l,N} \delta_{j,k}-\delta_{j,N} \delta_{k,N} \delta_{i,l}) \\
&\left\{ \tilde\chi^{i,j} , \tilde\chi^{k,l} \right\} = 2  \imath (N+1) \theta (\delta_{j,N} \delta_{k,N} \delta_{i,l} - \delta_{i,N} \delta_{l,N} \delta_{j,k}) \label{chichi} \end{align}
In addition we define:
\begin{align}
& \tilde\Delta^{(i,j),(k,l)} \equiv  \left\{ \tilde\phi^{i,j} , \tilde\chi^{k,l} \right\}  \label{phichi}
\end{align}
Where  the action of the kernel $\tilde\Delta$ is given by:
\begin{align}
&\tilde\Delta^{(i,j),(k,l)}C_{l,k} = - 2\left(\left[\left[C,a^\dagger\right],Z\right] + \left[\left[C,a\right],Z^\dagger\right] \right)_{ij} \label{DeltaC}\\
& C_{j,i} \tilde\Delta^{(i,j),(k,l)} = - 2\left(\left[a,\left[Z^\dagger,C\right]\right] + \left[a^\dagger,\left[Z,C\right]\right] \right)_{kl} \label{CDelta}
\end{align}

Equations (\ref{phiphi},\ref{chichi}) don't conform with (\ref{PhiIrreducible},\ref{chichic0}), where boundary terms ($i,j, k$ or $l$ equal to $N$) appear in the former, however there are redundancies: not all $\tilde\phi^{(a,b)}$ are independent, nor are the $\tilde\chi^{(a,b)}$ independent. In fact from the $N^2$ constraints (the $\tilde\phi$'s), only $N^2-N$ are independent.  The same may be said of the  $\tilde\chi'$s. To see this we can consider the relation $[X,Y]=i\ell_\theta^2(\mathds{1}-N|N><N|)$ as a linear relation determining the matrix $Y,$ given the matrix $X$.  There are  $N$  relations between these conditions. These relations may be written as $\tr ( X^n [X,Y])=0$,  for  $n=1,\dots.N$. 


 To choose a subset of $N^2 - N$ conditions, we may take $N^2-N$ conditions that we get for the case that the matrix $X$ is given by $X=\frac{a+a^\dagger}{2}$ (for such an $X$, a solution for $Y$ is given by $Y = \frac{a-a^\dagger}{2 \imath}$, and a correctly chosen set of conditions will remain  independent in a neighborhood of such a solution).   For such a choice of $X,$ the fact that $\tr ( X^n [X,Y])=0$  means that it is possible to predict the imaginary values of the elements on the $N$'s row and column of the matrix $[X,Y]$  given all other real and imaginary elements. This prompts us to  define a new set of conditions, $ {\phi}^{a,b}$, which are the same as $\tilde\phi^{a,b}$, but  with the elimination of the redundant elements. This set is given by:
 \begin{align}\label{tildephidefine}
  {\phi}^{a,b} = \left\{  \begin{array}{lr}
 \tilde\phi^{a,b} & a<N, b<N \\
 \mbox{Re}(\tilde\phi^{a,b}) & a=N \mbox{ or } b = N
 \end{array}
 \right.
 \end{align}

%

We  find a set of gauge conditions $ {\chi}$ which satisfies (\ref{chichic0}) and for which $ {\Delta}$  is non-singular for the solution $X=\frac{a+a^\dagger}{2}$,  $Y = \frac{a-a^\dagger}{2 \imath}$, again relying on the fact that the matrix will remain non-singular in a neighborhood of this solution. We choose:
 \begin{align}\label{tildechidefine}
  {\chi}^{a,b} = \left\{  \begin{array}{lr}
 \tilde\chi^{a,b} & a<N, b<N \\
 \mbox{Re}(\tilde\chi^{a,b}) & a=N \mbox{ or } b = N
 \end{array}
 \right.
 \end{align}

  Let us confirm that $ {\Delta}$ has no zero modes. The action of $ {\Delta}$ differs from the action of $\tilde\Delta$ in that $ {\Delta}$ acts only on matrices with real values on the $N$th row and column. To obtain the result of this action, one may compute the action of $\tilde\Delta$ on the same matrix and take the real part of the result  on the last row and column. We first look for matrices for which the  action of  $\tilde\Delta $  (according to (\ref{CDelta}))  produces only  values on the last row and column. Explicit calculation shows that the only matrices to satisfy this  are proportional to $a^n + a^{\dagger n}$ or to $\imath(a^n - a^{\dagger n})$. Since we may only act on matrices with real values on the last row and column, the former is  chosen. Finally, the result of the action of $\tilde\Delta$ on $a^n + a^{\dagger n}$ is confirmed to have real boundary values, showing that the action of $ {\Delta}$ is non-singular.

\section{Stochastic Quantization in the Coulomb gauge}

With this, the stochastic quantization equations, in imaginary physical and auxiliary times, may be written:
\begin{align} \label{StochasticQ}
\frac{d \hat Z}{d \tau}  =  \dot{\hat Z} + \imath\Omega \hat Z  + \imath  \left[  \hat Z , \lambda \right] + \left[ a , \nu \right] + d\Xi,
\end{align}
where $\lambda$ and $\nu$ are Lagrange multipliers, which are chosen such that the constraint and the gauge condition are satisfied after time differentiation. $\Xi$ is a complex matrix with white noise elements:
\begin{align}\label{NoiseCorrelations}
\< \Xi_{i,j}(t,\tau) \Xi^\dagger_{k,l} (t',\tau') \> = 2\ell_\hbar^2  \delta_{i,k} \delta_{j,l} \delta(t-t') \delta(\tau - \tau')
\end{align}
The classical analogue is solved by neglecting the noise ($\hbar \to 0$) and  the $\tau$ derivative.

The  Lagrange multipliers are determined by demanding that the gauge condition and the constraint continue to be obeyed with the passing of auxiliary time:
\begin{align}\label{ConstraintTimeDerivative}
 {\rm Re} \left( \left[ \frac{d \hat Z^\dagger}{d\tau} , \hat Z\right]\right) =0,
\end{align}
\begin{align}\label{GaugeTimeDerivative}
 {\rm Im} \left( \left[ \frac{d\hat Z^\dagger}{d\tau} , a \right]\right) =  \sum_i  \alpha_i { \rm Im} \left( |N\>\<i|\hat Z_{n,i}\right) ,
\end{align}
where $\alpha_i$ are arbitrary real constants, which arise since we  relax the conditions  ${\rm Im}[\hat Z^\dagger,a]=0$ by allowing some arbitrariness in the boundary elements  of ${\rm Im}[\hat Z^\dagger,a],$ as discussed above in this section. 

\section{Symmetry Algebera}
We now turn to review the integrability of the quantum model underlying the stochastic quantization procedure. We follow the method of reference [\onlinecite{Wadati:QCalogero:And:W}], which gives a  treatment of the hidden symmetries of the model, in a formalism which fits in nicely into the current approach.

 The conserved quantities, denoted by $I_n$ are given by:
\begin{align}\label{AllconservedQs}
I_n =(N+1) \left\< N | \mathcal{S}\left( \hat Z^{\dagger n} \hat Z^n \right) | N\right\>,
\end{align}
where $\mathcal{S}(\dots)$ denotes the symmetrized product, namely the average over all possible ordering of the operators which stand in place of  the ellipsis.   These operators all commute among themselves, which entails their conservation. The Hamiltonian is directly related to the first of these conserved quantities  as follows: $H = 2 m\Omega^2  I_1.$ Another set of operators which have a special role are given by $B_n $ :
\begin{align}
B_n =(N+1) \< N | \mathcal{S} Z^n| N \>
\end{align}
 These operators, rather than being conserved, obey the equation:
\begin{align}\label{BnCommutationWithH}
[H,B_n^\dagger] =\hbar \Omega n B_n^\dagger, \quad [H, B_n] =-\hbar \Omega n B_n,
\end{align}
demonstrating that they are spectrum generating operators.  The creation operators commute among themselves, as do the annihilation operators. To generate the eigenstates of the Hamiltonian, one first defines:
\begin{align}
h_j = \<j|\hat Z|N\>.
\end{align}
The Hamiltonian, is then easily shown to be given by:
\begin{align}
m \Omega^2 \sum_{i=0}^{N} h_i^\dagger h_i + \frac{\Omega( N+1)}{2} \left(\theta N+\hbar  \right).
\end{align}
The ground state is then annihilated by $h_i$ for all $i$, and excited sates are then created by acting with $B^\dagger_n$:
\begin{align}
|\{\lambda_k\}_{k=1}^l \> = \left(\prod_{k=1}^l B^\dagger_{\lambda_k}\right) |0\>,
\end{align}
the energy eigenvalue is then $E=\hbar \Omega \sum_k \lambda_k$, which is, of course, a consequence of (\ref{BnCommutationWithH}).

\section{Mode Expansion of Symmetry Algebra}

To reveal the conformal and the $W_\infty$ invariance, we consider now a hydrodynamic expansion of the conserved quantities. We expand $\hat Z$ as in (\ref{smallDeviation}) but now keeping higher order terms:
\begin{align}\label{Zexpansion}
\hat Z = a + V(a^\dagger) + F + \dots,
\end{align}
where $V$ is a small quantity and $F$ is regarded as second order in $V$. The ellipsis denotes yet higher order corrections. We expand $V(a^\dagger)$ in a power series:
\begin{align}\label{Vexpansion}
V(a^\dagger) = \sum_{n=1}^{N} c_{n}\alpha_{n+1} a^{\dagger n},
\end{align}
where
\begin{align}
c_{n}=\sqrt{\frac{ (N-n)!}{\ell^{2n}_\theta2^n(N+1)!}}\label{cnDef}
\end{align}
$V$ may thus be thought as a complex function defined by the power series $V(w) = \sum_{n=1}^N \alpha_{n+1} c_{n+1}w^n$ in which we may substitute $a^\dagger$. To make more precise what is meant by having $V$ small, we make the following assumption:
\begin{align}
\frac{\partial^n V(w)}{\partial w^n} \sim \frac{\ell_\theta}{L^{n+1}},
\end{align}
where $L$ is some macroscopic length scale over which the function $V$ varies. We assume $\ell_\theta \ll L$ and write the first two  conserved quantities to leading order in $\frac{\ell_\theta}{L}$, in terms of the modes $\alpha_n$, or rather in terms of the quantum operators $\hat{\alpha}_n$ corresponding to them.

Before writing the conserved quantities we first show that $\hat\alpha_n$ and $\hat\alpha^\dagger_n$ have standard commutation given by:
\begin{align}
[\hat{\alpha}_n, \hat{\alpha}^\dagger_n ] =2\ell_\hbar^2 n.
\end{align}
To derive this we write down the equation of stochastic equations of motion for $\alpha_n$ which may be derived from (\ref{StochasticQ})
by substituting (\ref{Zexpansion}) and (\ref{Vexpansion}). This leads to the following equation, which is shown to hold in Appendix \ref{FirstAppendix}:
\begin{align}
\frac{d \alpha_n }{d \tau} = \dot{\alpha}_n +\imath n\Omega \alpha_n  + \sqrt{n} d \xi_n,\label{SimpleHarmonics}
\end{align}
where $\< d\xi_n(t,\tau) d \xi_m (t',\tau') \> =  2\ell_\hbar^2  \delta(t-t') \delta(\tau- \tau') $.  We may now write the first conserved quantity , $I_1^{(0)}$ (which is proportional to the Hamiltonian). This is given by the simple expression:
\begin{align}
I_1^{(0)} =   \sum_n \hat\alpha^\dagger_n \hat\alpha_n + ( N+1) \left(\ell_\theta^2 N+\ell_\hbar^2  \right),\label{I1MainTex}
\end{align}
where the zero point energy is derived from comparison with the known exact spectrum.

The details of the derivation of (\ref{SimpleHarmonics})  is given in Appendix \ref{FirstAppendix}. The dynamics produced by $I_1^{(0)}$ is indeed trivial, however it is only the leading order approximation. All interesting physical effects are contained in the perturbations of this Hamiltonian. To describe the effect of the perturbation one must apply degenerate perturbation theory since $I_1^{(0)}$ has a  degeneracy which increases with energy level. Instead of directly diagonalizing the perturbation in the basis of degenerate states of $I_1^{(0)}$ one may make use of the existence of an infinite number of conserved quantities, in a procedure reminiscent of the one applied in  [\onlinecite{Wadati:Rodrigues}]. Indeed, the correct basis  must simultaneously diagonalize all $I_n^{(0)}$. In fact it is enough to consider $I_2^{(0)}$ which may be computed from (\ref{AllconservedQs}) to be given by (see Appendix \ref{SecondAppendix}):
\begin{align}
&I^{(0)}_2 =\ell_\theta\sum_{m,n} (\alpha^\dagger_{m+n}\alpha_m \alpha_n+h.c.) + \nonumber \\
&+ (\ell^2_\hbar-\ell^2_\theta)\sum_n n\alpha^\dagger_n \alpha_n +\ell_\theta^2 N \sum_n \alpha^\dagger_n \alpha_n
\label{I2Main}
\end{align}

It should be noted that we only explicitly derive here the $\hbar$ independent part of the Hamiltonian, while the $\hbar$ correction is inferred from knowledge of the exact spectrum (See Appendix \ref{SecondAppendix}). Indeed, the Hamiltonian (\ref{I2naught}) appeared already in the context of the Claogero model with no harmonic potential, where $\hat \alpha_n$ were related to the moments of the one dimensional density of the original Calogero-Sutherland-particles, rather than the moments of the two-dimensional fluid density, in the current approach. The convergence of the form of the Hamiltonians for these two different approaches, allows to deduce the $\hbar$ correction without need for lengthy computation. Although a more direct approach is, in principle, possible.

The expression, Eq. (\ref{I2Main}), for the second conserved quantity may also be recast as:
\begin{align}\label{I2naught}
I_2^{(0)} =\ell_\theta \sum_{n=1}^N \hat\alpha^{\dagger}_{n}  L_n+N\ell^2_\theta \sum_n \hat\alpha^{\dagger}_n \hat\alpha_n,
\end{align}
where
\begin{align}\label{LnModes}
L_n = \sum_{m=1}^N \hat\alpha^\dagger_m \hat\alpha_{n+m} + \sum_{m=1}^{n-1} \hat\alpha_m \hat\alpha_{n-m} + \left(\frac{\ell_\hbar^2}{\ell_\theta}-\ell_\theta\right)n\hat\alpha_n.
\end{align}

The problem of diagonalizing $I_2^{(0)}$ is directly related to conformal field theory, as shown in [\onlinecite{Mimachi:Yamada:Singular:Vectors:Jacks}]. This can be seen by considering that in the RHS of (\ref{I2naught}) there appears the Virasoro generator $L_n$, with central charge
\begin{align}\label{c}
c=1-6\left(\frac{\ell_\theta}{\ell_\hbar}- {\frac{\ell_\hbar}{\ell_\theta}}\right)^2,
\end{align}
This observation, made in [\onlinecite{Mimachi:Yamada:Singular:Vectors:Jacks}], leads to an expression for the eigenvectors of $I_2^{(0)}$ in terms of  Jack\cite{Jack:Of:Polynomial:Fame} symmetric functions\cite{macdonald:new:Class,stanley:1989:Jacks,macdonald:symmetric:polynomials} associated with a Young tableau, $2\leq\lambda_n \leq \lambda_{n-1}  \dots \leq \lambda_1, $ labeling the state (the $\lambda_i$'s are integers).   The first conserved quantities' eigenvalue  (the $I_1^{(0)}$ eigenvalue) of the state, to first order,  is then given by (\ref{Q1Eigenvalue}):
\begin{align}
E^{(0)}_\nu = \hbar \Omega \sum_i \lambda_i,
\end{align}

It is the relation between the problem of diagonlizing the Hamiltonian in a hydrodynamic expansion and the representation theory of the Virasoro algebra with $c$ given by (\ref{c}) that establishes the relation between quantum hydrodynamics of point vortices in two dimensions to conformal invariance, in the current settings. Classically, the conformal symmetry corresponds to the fact that the function  $V(w) $, describing the deformation of the droplet, and which must be analytic, remains analytic under a conformal transformation. This statement, a basic feature of analytic functions, becomes more structured quantum mechanically, in which case the symmetry algebra received a central extension, characterized by  a number $c$, which is written through the density of the fluid, $\ell_\theta^{-2}$ , in units of $\ell_\hbar^{-2}$.

\section{Conclusion}
We have studied two dimensional quantum hydrodynamics of point vortices in a model in which vorticity is directly related to density. The infinite symmetry algebra of the model studied leads to the description of eigenstates of the quantum system in terms of Jack  functions, which are intimately related\cite{Mimachi:Yamada:Singular:Vectors:Jacks,Awata:Shiraishi:Collective:Calogero:CFT,Awata:Singular:Vectors:Jacks,Wadati:Rodrigues,Wadati:QCalogero:And:W} to the representation theory of the Virasoro algebra and the $W_\infty$ algebra. The central charge for the Virasoro algebra is given in (\ref{c}).

Next, a hydrodynamic expansion is considered, in which only  collective motion is described. To zeroth order, the resulting dynamics  exhibiting only the solid rotation of the droplet. Despite the triviality of the dynamics, which is related only to the leading order in an expansion of one conserved quantity, namely,  the Hamiltonian,   the problem of diagonalizing all the conserved simultaneously, is a rich problem related to the representation theory of the conformal and a nonlinear $W_\infty$ algebras [\onlinecite{Awata:Singular:Vectors:Jacks}]. The central charge associated with the conformal algebra is given in (\ref{c}).  The solution of the problem is given in terms of  Jack polynomials\cite{Jack:Of:Polynomial:Fame}.

The next order in the expansion has non-trivial dynamics -- the actual hydodynamics of the droplet. To study it one may apply perturbation theory to the objects found to zeroth order in the expansion, and thus the problem of the dynamics of the droplet are also intimately related to representation theory of infinite dimensional symmetries.

The way in which conformal invariance is expressed in the current problem of quantum hydrodynamics, may shed light also on the problem of stochastic hydrodynamics and perhaps turbulence. Indeed, the stochastic quantization approach allows one to phrase our findings in terms of properties of a stochastic hydrodynamic process. In this respect it may be commented that although an unphysical  auxiliary time appears to interfere with the interpretation of stochastic quantization in terms of a classical stochastic process, this time is actually set to infinity in all computations, or alternatively phrased, the derivative with respect to this time is assumed to be zero in distribution, and thus is removed at the end of the computation.   

Importantly, the way in which conformal invariance appears is  not straightforward, but is related to a subtle connection between the solutions of such systems of stochastic equations, with the representation theory of the conformal algebra.  Developing this approach further will be the subject of future studies.

As a last remark we note that the fluid we study may be understood as being the Hall fluid. Indeed, this was the original motivation of Susskind\cite{104:Susskind:NC} to study the non-commutative Chern-Simons theory, which is the infinite version of the constraint matrix model,  introduced by Polychronakos\cite{Polychronakos:FQH:CS}, and which is also studied here. It should be noted, though, that the quantum Hall state is based on a topological classification of quantum many body states, such that the dynamics is ostensibly not automatically prescribed by the fact that a system is in a quantum Hall state. Nevertheless, certain physical situations may exist in which the hydrodynamics studied here may be the relevant dynamics of a quantum Hall system. Some hints of this scenario appeared in recent works of Wiegmann\cite{Wiegmann:FQHE:Edge,Wiegmann:Kirchoff:Larking} and Abanov\cite{Abanov:FQHE:Hydrodynamics}, however further work is needed in order to be able to flesh out such a possible relation.      

\section{Acknowledgements}
The author acknowledges extensive discussion and the collaborative effort of O. Agam, G. Falkovich, P. Wiegmann and A. Zamolodchikov. The author would also like to acknowledge discussion with A. Abanov, A. Altland and D. Bernard.
The author is grateful for the hospitality at the University of Cologne, where this work has been completed. This work has been supported by the Israel Science Foundation (Grant No. 852/11) and by the Binational Science Foundation (Grant No. 2010345).

\appendix

\section{Computation of first  Hamiltonian \label{FirstAppendix}}
It will prove convenient in the following to introduce an auxiliary-physical time, $T$, as follows:
\begin{align}
T = 2(\tau  + t ), \quad 
\end{align}

To compute the Hamiltonian, it is useful to  define the quantum part of the velocity as the part  proportional to the white noise, and the classical part as the deterministic part of the velocity. Explicitly we define the quantum part as:
\begin{align}\label{vQ}
v^Q = d\Xi + \imath [\hat Z, \lambda^Q] + [a,\nu^Q]
\end{align}
where $\lambda^Q$ and $\nu^Q$ must solve:
\begin{align}
\mbox{Re} \left([v^{Q},\hat Z^\dagger ]\right)=0 \label{ConstaintAsRe}\\
\mbox{Im} \left([v^{Q},a^\dagger ]\right)=0, \label{GaugeAsIm}
\end{align}
and the classical part as:
\begin{align}\label{vcl}
v^{cl}= \imath \Omega \hat Z + \imath [\hat Z, \lambda^{cl}] + [a,\nu^{cl}]
\end{align}
where the classical and quantum parts  of the Lagrange multipliers solve:
\begin{align}
\mbox{Re} \left([v^{cl,Q },\hat Z^\dagger ]\right)=0\\
\mbox{Im} \left([v^{cl,Q},a^\dagger ]\right)=0,
\end{align}
respectively. Finally, the velocity in auxiliary time is just the sum of the two velocities, and the Lagrange multipliers are the sum of the two contributions:
\begin{align}
\frac{d\hat Z}{dT} = v^{cl} + v^{Q}, \quad \lambda=  \lambda^{cl} + \lambda^Q, \quad \nu = \nu^{cl} + \nu^{Q}.
\end{align}

We solve the equations by iteration assuming  the $\alpha_n$'s are  small. The order of magnitude of the noise is assumed to be such that the square of the noise if of the order of $\alpha_n$. We start with $v^{cl(0)}$, the zeroth order of the classical part of the velocity in auxiliary time. The quantity, $v^{cl(0)}$ features the zeroth order term of the Lagrange multipliers, $\lambda^{cl(0)}$ and $\nu^{cl(0)}$, namely the $\alpha_n$ independent part.
These satisfy:\begin{align}
\left[a^\dagger,\imath \Omega a + \imath [ a , \lambda^{cl(0)} + i \nu ^{cl(0)}  \right] =0.
\end{align}
This equation is solved by taking $\lambda^{cl(0)} =- \frac{\Omega^2 a^\dagger a}{2m\theta}$ and $\nu^{cl(0)} =0$, which leads to
\begin{align}\label{vcl0}
v^{cl(0)}=0
\end{align}

We next compute the first order correction to $v^{cl}$:
\begin{align}
v^{cl(1)} = \imath[V(a^\dagger),\lambda^{cl(0)}] +\imath\Omega V +\imath[a,\Phi ^{cl(1)}],
\end{align}
where we have introduced the notation:
\begin{align}
\Phi = \lambda+ \imath \nu,
\end{align}
which we shall now use in the sequel. Now, $\Phi^{cl(1)}$ is determined by combining Eqs. (\ref{ConstaintAsRe}, \ref{GaugeAsIm}), which yields:
\begin{align}
\imath [a^\dagger, v^{cl(1)}]=0,
\end{align}
solved trivially by $\lambda^{cl(1)} = \nu^{cl(1)} =0$. We obtain:
\begin{align}\label{vcl1}
v^{cl(1)} =  \sum_n \left(\dot{\alpha}_{n+1} -\imath\Omega(n+1)\alpha_{n+1} \right)c_{n} a^{\dagger n},
\end{align}
where $c_n$ is defined in (\ref{cnDef}).

We proceed to compute the quantum part of the velocity in auxiliary time, $v^Q$, which contains no zeroth order part.  First we note that $v^{Q(1)}$, defined as:
\begin{align}\label{vQ1def}
v^{Q(1)} = \imath [a,\Phi^{Q(1)}] + d\Xi
\end{align}
 satisfies:
\begin{align}
[a^\dagger , v^{Q(1)} ]  =0,
\end{align}
which, in turn, implies that  $v^{Q(1)}$ is a power series in $a^\dagger$:
\begin{align}\label{vQ1PowerSum}
v^{Q(1)} = \sum_n d\xi_{n+1}c_n a^{\dagger n}.
\end{align}
To compute the coefficients $d\xi_n$, we note that
\begin{align}\label{traceAdaggerNaN}
\tr(a^{\dagger n } a^m ) = \frac{1 }{(n+1)c_{n}^2} \delta_{n,m},
\end{align}
thus hitting (\ref{vQ1PowerSum}) with $a^n$ and taking a trace on the one hand and hitting (\ref{vQ}) with $a^n$ and taking the trace on the other, one obtains:
\begin{align}\label{dxindef}
d\xi_{n+1} =(n+1)c_{n}\tr\left(a^n d \Xi\right).
\end{align}
Computing the cross-correlations of these noises is straightforward using this equation and (\ref{NoiseCorrelations}). The result is:
\begin{align}
\<d\xi_n d\xi^*_m\>  =\frac{2\hbar n}{m \Omega}  \delta_{n,m} \delta(t-t')\delta(\tau-\tau')
\end{align}
Combining (\ref{vcl0}, \ref{vcl1}, \ref{vQ1PowerSum}) and (\ref{Vexpansion}), one obtains:
\begin{align}
\frac{d\alpha_n}{  d T } =-\imath  n\Omega\alpha_n  + d\xi_n
\end{align}
These are the equation in stochastic quantization obtained for the quantum system of harmonic oscillators for the Hamiltonian $H=m\Omega^2 I_1$ with :
\begin{align}
I_1 =\sum_n \hat\alpha_n^\dagger \hat\alpha_n +C_1, \quad [\hat\alpha_n , \hat \alpha^\dagger_m ]  =n\ell_\hbar^2  \delta_{n,m},\label{I1Appendix}
\end{align}
where $C_1$ is a constant which is not fixed by our considerations. To compute the constant consider  $I_1,$ given in Eq. (\ref{AllconservedQs}), was shown in Ref. [\onlinecite{Wadati:QCalogero:And:W}] to give simply the Calogero Hamiltonianm where it was also confirmed to give the same spectrum as that found in the  original work of Calogero \cite{Calogero:1971}. The spectrum of $I_1$ is then :
\begin{align}
Q_1 =\ell_\theta^2 \left(\sum_i i \lambda_i +( N+1) \left( N+\frac{\ell_\hbar^2}{\ell_\theta^2}  \right)\right),\label{Q1Eigenvalue}
\end{align}
for any set of integers $2\leq \lambda_n \leq \lambda_{n-1} \dots \leq \lambda_1$ (a partition). This spectrum coincides with the spectrum which may be easily derived for  (\ref{I1Appendix}) if $C_1$ is chosen as follows, $C_1 = \frac{( N+1)\left( \theta N+\hbar  \right)}{m\Omega} $.  With this choice of $C_1$  we get  (\ref{I1MainTex}).

\section{Computation of second Hamiltonian\label{SecondAppendix}}

We assume now the evolution:
\begin{align}
\frac{d\hat Z}{dT} =  - \imath\Omega \ell_\theta^{-2} \mathcal{S}\left[ \hat Z^\dagger \hat Z \hat Z \right] + \imath [\lambda, \hat Z]+[\nu,a] + d\Xi
\end{align}
which according to [\onlinecite{Wadati:QCalogero:And:W}] is the stochastic quantized version of the evolution\cite{} generated by $\frac{m^2 \Omega^3}{\theta} I_2.$ We remind that $\mathcal S[\dots]$ denotes the symmetrized product. We expand this equation in modes. We shall only deal with the classical part by ignoring the noise and assuming $\nu^Q=\lambda^Q=0$, restoring quantum corrections to the Hamiltonian at the end of the calculation, based on the known spectrum of $I_2$.

 The zeroth order equation gives, $v^{cl(0)}=0$ and:
\begin{align}\label{0thOrder2ndH}
[\lambda^{cl(0)}, a] =\ell_\theta^{-2}\mathcal{S}\left[a^\dagger a a \right] .
\end{align}
The first order equation gives:
\begin{align}\label{2ndHvcl1}
   v^{cl(1)} &= - \imath\Omega\ell_\theta^{-2}\mathcal{S}\left[  2a^\dagger V  a+V^{\dagger  } aa\right] + \nonumber \\
 &+\imath [
\lambda^{cl(0)} , V] + \imath [\lambda^{cl(1)},a],
\end{align}
with the condition:
\begin{align}
[v^{cl(1)} , a^\dagger ] =0.
\end{align}
Making use of the expansion (\ref{Vexpansion}), one obtains, to leading order:
\begin{align}
\frac{d\alpha_{n+1}}{dT}  =\imath c_n (n+1) \tr\left[ a^{n}\left([\lambda^{cl(0)} , V]-2\ell_\theta^{-2}\mathcal{S}\left[  a^\dagger V  a \right] \right)  \right]
\end{align}
The traces may be computed as follows. First simple algebra using the explicit expression for $a,$ Eq. (\ref{aDef}), gives:
\begin{align}\label{2ndH1stTr}
\tr \left[a^n \mathcal{S}(a^\dagger V a) \right] = \alpha_{n+1} \left(\frac{c_{n}}{(n+2)c^2_{n+1}} +\frac{ 2\ell_\theta^2}{6 c_n}\right).
\end{align}
The other trace we need to compute is given by:
\begin{align}\label{2ndH2ndTr}
&\tr\left[a^{\dagger n} \imath [ \lambda^{Q(0)} , V^\dagger]\right]^* =\\
&=\alpha_{n+1} c_n \sum_{j=0}^{n-1}\tr\left[\imath a^{\dagger n} a^j \mathcal{S}(a^\dagger aa)a^{n-j-1}\right]^*, \nonumber
\end{align}
where (\ref{0thOrder2ndH}) have been used. The trace under the summation sign on the RHS  may  be computed explicitly:
\begin{align}\label{2ndH3rdTr}
&\tr\left[\imath a^{\dagger n} a^j \mathcal{S}(a^\dagger aa)a^{n-j-1}\right] = \\
&=\imath \left[ \frac{1}{c^2_{n+1}(n+2)} + \frac{2\ell_\theta^2( j+1)}{c_n^2 (n+1)} - \frac{2\ell_\theta^2\delta_{j,n-1}} {3 c_{n}^2 } \right] \nonumber
\end{align}
Combining (\ref{2ndH1stTr}, \ref{2ndH2ndTr}) and (\ref{2ndH3rdTr}) we obtain:
\begin{align}
\frac{d\alpha_n}{\Omega dT} = -\imath n \left[2N-   n \right] \alpha_n.
\end{align}
We proceed to compute the higher order terms in the equation of motion for $\alpha_n$, namely those terms which are quadratic in the $\alpha_m$'s. The classical velocity to second order is given by:
\begin{align}\label{vcl2.2ndH}
v^{cl(2)} &=\Omega      \ell_\theta^{-2} \mathcal{S}(VVa^\dagger)+ \imath[\lambda^{cl(1)},V]\\
&+\Omega\ell_\theta^{-2}\mathcal{S}(2V^\dagger V a)+\imath [\lambda^{cl(0)},F]+\imath [\lambda^{cl(2)},a]-\dot{F}^{cl(2)} \nonumber
\end{align}
At this point we only need the zeroth order terms in a $1/N$. As a consequence,  it is fairly easy to compute the contribution of the first three terms in the first line of (\ref{vcl2.2ndH}). Furthermore, we may concentrate on computing only bilinear terms in $\alpha_m$, such as $\alpha_m \alpha_{n-m-1}$, ignoring sesquilinear combinations. such as $\alpha^*_m \alpha_{n+m+1}$. The latter may be finally restored by considering that $I_2$ must be  Hermitian. Such bilinear forms originate only from the first three terms in  (\ref{vcl2.2ndH}). The traces compute as follows. First, one may compute rather easily:
\begin{align}
&(n+1)c_{n}\tr\left[a^{ n}\mathcal{S}(VVa^\dagger) \right] =
\sqrt{2} \ell_\theta\sum_m \alpha_{m+1}\alpha_{n-m},
\end{align}
which is correct to leading order in $1/N$. Continuing with the next term we have:
\begin{align}
&\tr\left[a^n\imath[\lambda^{cl(1)},V]\right] = \\
& =\sum_m c_m\alpha_{m+1} \tr\left[a^n a^{\dagger j }\imath[\lambda^{cl(1)},a^\dagger] a^{\dagger (m-1-j)}\right]\nonumber.
\end{align}
Using (\ref{2ndHvcl1}) and retaining only bilinear terms one obtains:
\begin{align}
&(n+1)c_n \tr\left[a^n\imath[\lambda^{cl(1)},V]\right] = \\
& =\imath\frac{\ell_\theta}{\sqrt{2}}  (n-1)\alpha_{m+1}\alpha_{n-m} + \dots,
\end{align}
where the ellipsis here denotes sesquilinear terms and higher order corrections in $1/N$. As a result we obtain the bilinear term in the dynamics:
\begin{align}
\frac{d\alpha_{n}}{\Omega dT} =  \frac{\imath n}{\ell_\theta}\sum_m \alpha_{m} \alpha_{n-m} + \dots.
\end{align}
This term is generated by a term in the Hamiltonian of the form $\alpha^\dagger_n \alpha_m \alpha_{n-m-1}$, to which we must add the Hermitian conjugate $\alpha^\dagger_{n-m-1} \alpha^\dagger_m \alpha_n$. Together we have the following classical dynamics:
\begin{align}
 \frac{d\alpha_n}{\Omega dT} &=\frac{\imath n}{\ell_\theta} \left[\sum_m \alpha_m \alpha_{n-m} + 2{ } \sum_m \alpha^*_m \alpha_{n+m} \right]+ \nonumber\\
&+2\imath n\left[(2N+1)-n)\right] \alpha_n.
\end{align}
The  dynamics described by the latter equation are generated by  a Hamiltonian,$\frac{m^2 \Omega^3}{\theta}  I_2$, with $I_2$ given by:
\begin{align}
&I_2 =\ell_\theta\sum_{n,m} \left(\alpha^\dagger_n \alpha_m \alpha_{n-m} +h.c.\right) - \nonumber \\ &- (\ell_\theta^2+C_2\ell_\hbar^2) n  \alpha^\dagger_n \alpha_n +(N\ell_\theta^2+C_3\ell_\hbar^2)  I_1+\ell_\hbar^4 C_4,     \label{SecondConservedExplicit}
\end{align}
where by introducing $C_2$, $C_3$, $C_4$ we have added the only possible term that could arise from a regular expansion in $\ell_\hbar^2$ and that is no smaller in powers of  $\frac{\ell_\theta}{L}$ than the  rest of the terms in $I_2$. We only consider terms that commute with $I_1$.  The spectrum of the second conserved quantity is known from Ref. [\onlinecite{Wadati:Rodrigues}]. In contrast to  Ref. [\onlinecite{Wadati:Rodrigues}], we write the result in terms of the  dual Young diagram (such as is done, e.g., in Ref. [\onlinecite{Awata:Shiraishi:Collective:Calogero:CFT}]) as :
\begin{align}
Q_2 =\ell_\theta^2\sum_i \left[\left(N+\frac{\ell_\hbar^2}{\ell_\theta^2}(1-2i)\right) \lambda_i -\lambda_i^2  \right],
\end{align}
 for  partitions $2\leq \lambda_n \leq \lambda_{n-1} \dots \leq \lambda_1$. This form of the spectrum coincides with a second conserved quantity  of the form (\ref{SecondConservedExplicit}) for those values of $C_2, C_3$ and $C_4$, that prompt (\ref{I2Main}).

\bibliographystyle{nar}
first three terms in\bibliography{mybib}

\end{document}